\documentclass[a4paper,12pt]{article} 
\usepackage{graphicx} 
\usepackage{graphics} 
\usepackage{amsmath} 
\usepackage{amsfonts}  
\usepackage{amssymb} 

\usepackage[colorlinks=true, pdfstartview=FitV, linkcolor=blue,  citecolor=blue, urlcolor=blue]{hyperref}  

\usepackage{supertabular,booktabs}

\usepackage{amsmath}   

\def\l{\left}
\def\r{\right}
\def\be{\begin{equation}}
\def\ee{\end{equation}}
\def\beq{\begin{equation}}
\def\eeq{\end{equation}}

\begin{document}

\title{A catalog of UGC isolated galaxy pairs with accurate radial velocities}
\author{Pierre Chamaraux$^1$ and Laurent Nottale$^2$\\
 {\small $^1$ GEPI, CNRS, Paris Observatory, 92195 Meudon CEDEX, France}\\ 
 {\small pierre.chamaraux@obspm.fr}\\
{\small $^2$ LUTH, UMR CNRS 8102,  Paris Observatory, 92195 Meudon CEDEX, France}\\
{\small laurent.nottale@obspm.fr}}
\maketitle

\begin{abstract}
The present paper is devoted to the construction of a catalog of isolated galaxy pairs from the Uppsala Galaxy Catalog (UGC), using accurate radial velocities. The UGC lists 12921 galaxies to $\delta> -2^\circ 30'$ and is complete to an apparent diameter of 1 arcmin. 
       The criteria used to define the isolated galaxy pairs are the following: 1) Velocity criterion: radial velocity difference between the members lower than 500  km s$^{-1}$; 2) Interdistance criterion: projected distance between the members smaller than 1 Mpc; 3) Reciprocity criterion: each member is the closest galaxy to the other one, which excludes multiplets; 4) Isolation information: the catalog lists the ratio $\rho$ between the projected distance to the closest UGC galaxy (having a velocity difference smaller than 500 km.s$^{-1}$) and the pair members interdistance, thus allowing one to choose any isolation criterion (beyond the chosen limit $\rho>2.5$). In addition, we have accounted for the small diameter bias by searching for CGCG galaxies in the pair environment and used the same isolation criterion.
       A peculiar investigation has allowed to gather very accurate radial velocities for pair members, from high quality HI and optical measurements (median uncertainty on velocity differences 10 km s$^{-1}$). Our final catalog contains 1005 galaxy pairs with $\rho>2.5$, of which 509 have $\rho>5$ ($50 \%$ of the pairs, i.e. $8 \%$ of the UGC galaxies)  and 273 are highly isolated with $\rho>10$ ($27 \%$ of the pairs, i.e. $4 \%$ of the UGC galaxies).
       Then we give some global properties of the pair catalog. We display the histograms of the radial velocity differences between the pair members and of their projected interdistances (median 0.29 Mpc). Finally, we provide an estimate of the contamination by cosmological false ``pairs", which is about 10\% up to a velocity difference of 380 km s$^{-1}$, beyond which all pairs are probably false.
\end{abstract}

\section{Introduction}

      The isolated galaxy pairs represent the simplest gravitational systems of galaxies. Their dynamical study is especially fruitful to estimate the masses and mass-luminosity ratios of their members, and possibly to check the presence of massive haloes and of dark matter (see for instance the basic works by Peterson (1979) \cite{Peterson1979} and by Chengalur et al. (1996) \cite{Chengalur1996}). Such studies need a large catalog of those galaxy pairs, with accurate radial velocities of their members. A pioneering catalog of isolated galaxy pairs has been devised by Karachentsev (1972) \cite{Karachentsev1972}; it lists 603 pairs north of the declination $\delta=-3 ^\circ$ and it has been used in several studies. Since then, other catalogs of galaxy pairs have been made available, for instance by Soares et al. (1995) \cite{Soares1995} which completes Karachentsev'one in the Southern hemisphere, and especially by Karachentsev and Makarov (2008) \cite{Karachentsev2008}, which gathers 509 bound pairs in the Local Supercluster with radial velocities $V_r<3500$ km s$^{-1}$.
      
      Recently, several works have been devoted to the study of the enhancement of star formation activity in pairs of galaxies due to gravitational interaction, resulting in new pairs catalogs. The most important number of pairs (nearly 13 000) have been used for this purpose after extraction from SDSS-DR2 and 2dFGRS by Alonso et al. (2006) \cite{Alonso2006}.
      
     In the present paper, we construct a catalog of isolated galaxy pairs from the UGC (Uppsala General Catalog of Galaxies, 1973) \cite{UGC}. The UGC covers the Northern Sky at $\delta >-2^\circ 30'$, and it is complete in apparent diameters to 1 arcmin. Since the linear diameter function of galaxies is known (Hudson \& Lynden-Bell (1991) \cite{Hudson1991}), one can consequently make appropriate corrections for the loss of galaxies with the distance in our catalog when necessary.
     
     Our main purpose is a dynamical statistical study of the isolated pairs of galaxies, which will be developped in next papers. A specific study will be devoted to the statistical determination of the actual velocity differences and the actual distances between the members of the pairs, which are needed in order  to understand their dynamics.
     
     The paper is organized as follows: section \ref{sec2} is devoted to a short description of the UGC and to its diameter function; in section \ref{sec3}, we present and apply the criteria used in order to define the isolated galaxy pairs; the gathering of accurate radial velocities of the members is detailed in section \ref{sec4}; the final catalog of 1009 galaxy pairs is given in section \ref{sec5}, some of its general statistical properties and corrections for cosmological false pairs are discussed in section \ref{sec6} and we conclude in section \ref{sec7}.

\section{Basis of the pair catalog: a brief reminder on Nilson's UGC}
\label{sec2}

\subsection{Uppsala General Catalogue}

The Uppsala General Catalogue of Galaxies (UGC) is an essentially complete catalog of galaxies to a limiting diameter of 1.0 arcminute and/or to a limiting apparent magnitude of 14.5 on the blue prints of the Palomar Observatory Sky Survey (POSS), constructed by P. Nilson \cite{UGC}. Coverage is limited to the sky north of declination $- 2.5$ degrees. The catalog contains 12921 galaxies.

However, few radial velocities of galaxies were available at the date of its completion (1973). Therefore, the UGC should be completed by an independent research of radial velocities, which we have done using Hyperleda (see the following). At the date of completion of the present pair catalog, velocities were known for 12141 galaxies, i.e. for 94 \% of the UGC galaxies (see Fig.~\ref{DiamFunct}).

\subsection{Diameter function for the UGC}
\label{diamfunct}

Using our UGC galaxy velocity distribution, we have the opportunity to check the Hudson and Lynden-Bell diameter function \cite{Hudson1991} and possibly improve it. Indeed, this diameter function is defined by:
\beq
\Phi_e (D) dD= \Phi_e^* \, \exp (-D/D^*) \, d(D/D^*)
\label{diam}
\eeq
where $ \Phi_e (D) dD$ is the number of galaxies by volume unit with diameter in the range $[D,D+dD]$, $\Phi_e^*$ is the density of UGC galaxies by volume unit (for all diameters) and $D^*$ is a characteristical diameter.

A galaxy at a distance $r$ is included in the UGC if its apparent diameter is larger than 1 arcmin, i.e. if its linear diameter $D_r>\pi r/10.8$ (for $D_r$ in kpc and $r$ in Mpc). Therefore, from Eq.~(\ref{diam}), the number $dN(r)$ of UGC galaxies between the spheres of radii $r$ and $r+dr$ is given for an homogeneous distribution by 
\beq
dN(r) = 4 \pi f \Phi_e^* \exp (-D_r/D^*) r^2 dr,
\eeq
where $f$ is the fraction of the sky surface covered by the catalog ($f=0.42$ for the UGC, accounting for the obscuration zone from our Galaxy).

If we transform the distance $r$ into radial velocity $v$ through the Hubble law, we obtain
\beq
dN(v)= 4 \pi f \,  \frac{\Phi_e^*}{H_0^3} \exp(-v/v^*)   v^2  dv
\label{dNv}
\eeq
for the number of UGC galaxies in the interval $v$ to $v+dv$, where $v$ is the Hubble velocity at distance $r$ and $v^*$ is the velocity of a galaxy of diameter $D^*$ seen with an apparent diameter of 1 arcmin.

\begin{figure}[!ht]
\begin{center}
\includegraphics[width=12cm]{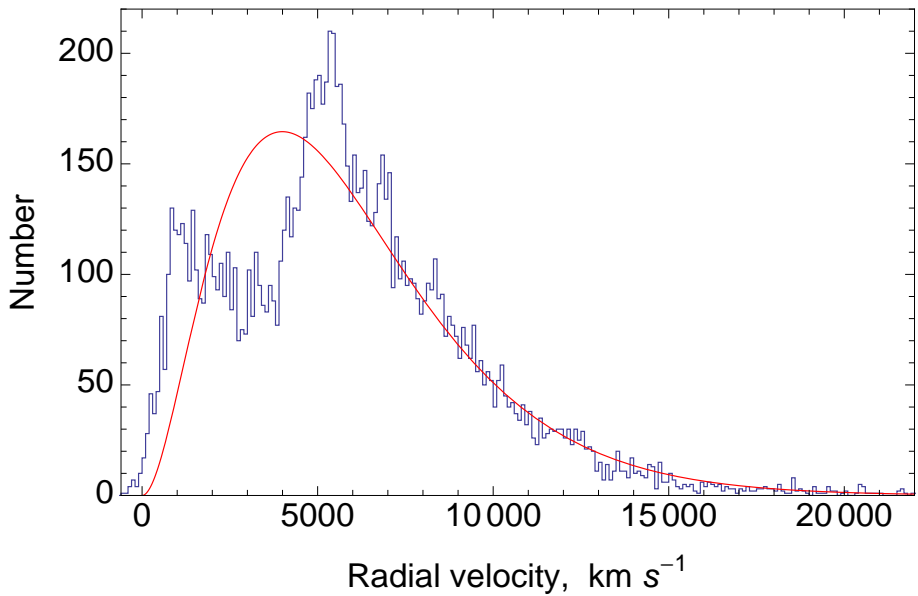}
\caption{{\small  Histogram of UGC galaxies radial velocities (blue line), plotted with a bin of 100 km s$^{-1}$; the ordinate is the number of galaxies per bin. The red curve is a fit of the distant part of this distribution ($v>7500$ km s$^{-1}$) through the diameter function (assuming an homogeneous distribution). } }
\label{DiamFunct}
\end{center}
\end{figure}

We have least-square fitted with this law the tail of the new velocity data  ($v>7500$ km s$^{-1}$, in order to remove nearby fluctuations due to clustering) and we have obtained (on $\approx 3300$ galaxies)
$v^*=1997 \pm 85$ km s$^{-1}$ 
 (see Fig.~\ref{DiamFunct}).  By integration, we obtain $\Phi_e^*=N H_0^3/(8 \pi f v^{*3})$, with $N=12141$ (the number of galaxies having available radial velocities). This leads to  a value of the catalog density $\Phi_e^*=(0.144 \pm 0.018) \; h^3$ Mpc$^{-3}$. 
 
 These results agree within error bars with the Hudson and Lynden-Bell  \cite{Hudson1991} values $v_0=2200 \pm 150$  km s$^{-1}$ and $\Phi_e^*=(0.15 \pm 0.03) \; h^3$ Mpc$^{-3}$ and improve them. Note that  the Hudson-Lynden-Bell value of $v_0$ was a fit made with a smaller number of radial velocities $N= 1510$ available at that time. The error they quote, 150  km s$^{-1}$, takes into account systematic errors specific to the UGC apparent diameters.

\subsection{Nilson's galaxy pairs}

The UGC also includes, in the description given of each galaxy and its surrounding area, a mention (pw : ``pair with") of its possible membership to a pair with another galaxy of the catalog. From this information, a Nilson list of 596 pairs has been extracted \cite{NottaleNilsonpair}. However, Nilson used  qualitative criteria (by-eye estimation on the POSS) for the definition of isolated pairs, and few radial velocities were available at the time of completion of the UGC. 

In the framework of our construction of the new pair catalog based on quantitative criteria and accurate radial velocities for almost all galaxies of the UGC, we have put to the test the validity of Nilson's pairs. We find that 395 among the 596 Nilson pairs have a radial velocity difference between their members smaller than 500 km s$^{-1}$, i.e. 66 \% of the pairs, and 362 smaller than 350 km s$^{-1}$ (61 \% of the Nilson pair list).  

Now, if we compare with our new catalog (see the following), we find 348 pairs among the 362 which satisfies our other quantitative criteria (of projected separation of the members and of relative isolation), i.e. 96 \% of the pairs. However, the total number of pairs in the present new UGC pair catalog is 1246, i.e. 3.5 times the number of real (non-cosmological) pairs in Nilson's list.

We conclude that the main problem in the constitution of Nilson's pair list was the lack of radial velocities; moreover, his qualitative isolation and separation criteria are validated by our quantitative ones, but they appear as more stringent as our's, since only 30~\% of our pairs are in the Nilson list.

\section{Definition of isolated galaxy pairs and method}
\label{sec3}

A galaxy pair is characterized by 6 variables, the three coordinates of the galaxies interdistance $(x, \,y,\, z)$, and the three coordinates of the velocity difference between its members, $(v_x,\, v_y, \, v_z)$. However, only three among these 6 parameters are observationally available. Namely, (assuming the $z$-axis  is oriented from the observer and therefore $(x, \,y)$ is in the plane of the sky), only $x,\,y$ and the radial velocity difference $v_z$ are observable. From the $x$ and $y$ coordinates, we can compute  the interdistance between the pair members projected on the sky plane, $r_p= \sqrt{x^2+y^2}$. In practice, this projected interdistance is derived from the observed angle on the sky $\theta$ between the galaxies and the pair distance $D$, i.e., $r_p=D \, \theta$ (since $\theta\ll 1$ except for a few close-by pairs).

Due to these observational constraints, we have to choose our criteria of pair definition from these two parameters, $v_z$ and $r_p$. We have made the choice to take not too constraining limits, in order not to miss possible real pairs having remote members. Doing that, we expect to introduce false (cosmological) pairs, which will be accounted for and excluded in the subsequent analysis.

Our criteria of selection are:

\begin{itemize}

\item Small interdistance: projected separation $r_p < r_{\rm lim}$, where we have taken $r_{\rm lim}=1$ Mpc (recall that the interdistance between the Milky Way and M31 is 0.7 Mpc).

\item Small velocity difference: we require a radial velocity difference $v_z<V_{\rm lim}$, where we have chosen the large value $V_{\rm lim}=500$ km/s (knowing that we shall subsequently correct for the false pairs, the ``members" of which have large velocity differences mainly due  to Hubble expansion).

\item Reciprocity criterion: we require that, if  the closest galaxy to a galaxy A  is B, then A is also the closest to B. This is a first isolation criterion, which also allows to exclude multiplets.

\item Relative isolation information:  $d_3$ being the projected distance of the closest galaxy from the UGC to the pair (also such that $|\Delta V_r|<500$~km.s$^{-1}$ with respect to the pair center), we have computed the ratio $\rho=d_3/r_p$ for all pairs satisfying the three first criteria. The final pair catalog lists all pairs such that $\rho>\rho_l=2.5$ and includes the value of $\rho$ for each pair. Therefore, sub-catalogs of weakly isolated pairs ($\rho>2.5$), isolated pairs (e.g. $\rho>5$) and highly isolated pairs (e.g.  $\rho>10$) are included in the pair catalog. This allows the user to choose his own isolation criterion. Note that the external force exerted on each member by another galaxy is on average less than $\approx 1/\rho^2$ that exerted by the other member, allowing the isolated pairs to be dynamically independant and physical.

\item Additional isolation criterion: the previous isolation criterion is biased at the low diameter limit (1 arcmin) of the catalog, since possible smaller galaxies present in the pair environment will not be members of the UGC and will be missed. Therefore we have decided to use a deeper catalog, Zwicky et al.'s Catalog of Galaxies and Clusters of Galaxies (CGCG, \cite{CGCG}) for searching galaxies in the pair environment. The magnitude limit of the CGCG ($m<15.7$) corresponds to a mean limiting UGC diameter (blue major axis) of $\approx 0.8$. Finally, we have kept all pairs such that the diameters of both members are $d \geq1.2$ arcmin and for the other pairs, we have used the same isolation parameter $\rho_Z=d_{3Z}/r_p$  (as previously done on UGC galaxies) on CGCG galaxies having a radial velocity difference from the pair center $|\Delta V_r|<380$~km.s$^{-1}$. Then all pairs such that $\rho_Z<2.5$ have been excluded from the final catalog.
We have also excluded pairs with $d \geq1.2$ but with $\rho_Z<0.5$.

\end{itemize}

\begin{figure}[!ht]
\begin{center}
\includegraphics[width=12cm]{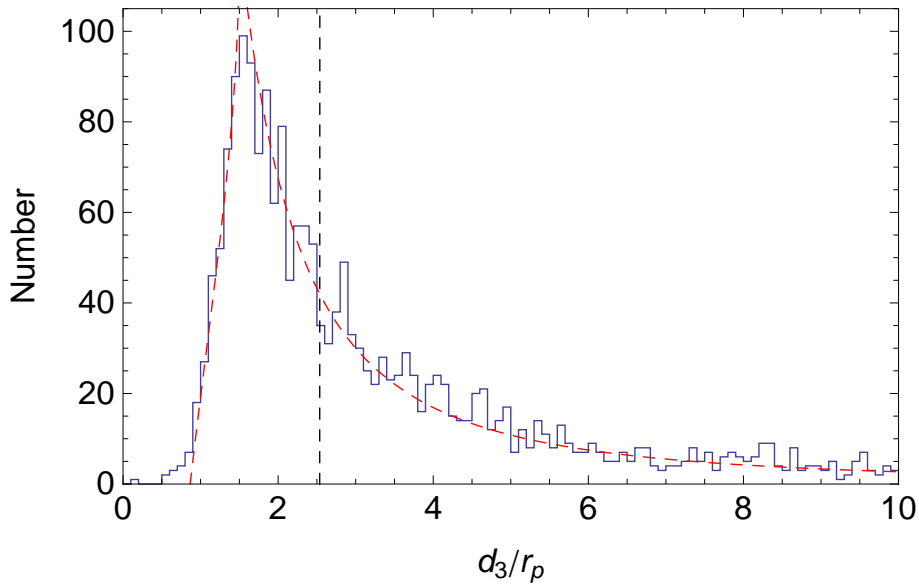}
\caption{{\small  Observed distribution of $\rho=d_3/r_p$ for all 2280 reciprocal pairs; $r_p$ is the projected interdistance between members of a pair; $d_3$ is the projected distance between the center of the pair and the nearest UGC galaxy. The limit of our pair catalog ($\rho>\rho_l=2.5$) is shown (vertical dashed line). The theoretically expected inner depletion due to the partial isolation constraint involved by the reciprocity criterion (see Fig.~\ref{cercles} and Eq.~\ref{eq4}) is displayed as a red dashed line. A $\rho^{-2}$ fit of the outer part is displayed as a dashed red curve.} }
\label{rho3}
\end{center}
\end{figure}

\begin{figure}[!ht]
\begin{center}
\includegraphics[width=6cm]{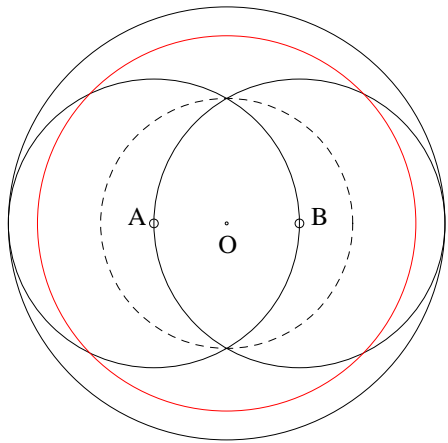}
\caption{{\small Illustration of the isolation constraint included in the reciprocity criterion. This criterion excludes the presence of a third galaxy in the two circles of radius $r_p$ centred on each member (A and B) of the pair of center O. This exclusion is complete from $\rho=0$ to $\rho= \sqrt{3}/2\approx 0.866$ (dashed circle), then partial at larger values (red circle) and it finally ends at $\rho= 3/2$. } }
\label{cercles}
\end{center}
\end{figure}

The method used to apply these selection criteria is the following. 

We start from a numerical version of the UGC catalog \cite{UGC}, which lists 12939 galaxies. We have completed it by radial velocities from the HyperLeda extragalactic database \cite{HyperLeda,Makarov2014}. We take the velocity $V_{\rm lg}$ corrected to the centroid of the Local Group. There were 12141 available radial velocities at the time of the completion of this version of the UGC. The distances $D$ of the galaxies were taken from the HyperLeda ``mod0" (redshift-independent modulus), which is an estimate of the luminosity-distance independently of the redshift. Otherwise, they were derived from the corrected redshift $V_{\rm lg}$ through the Hubble relation $D=V_{\rm lg}/H_0$, using $H_0=70$ km s$^{-1}$ Mpc$^{-1}$. 

We have written a Mathematica program searching around each galaxy of the UGC catalog all close-by galaxies such that their projected distance on the sky is $d<1$ Mpc, both for $d=\theta D_1$ and $d=\theta D_2$, $D_1$ being the cosmological distance of the reference galaxy and $D_2$ the cosmological distance of each of the selected galaxies. Then we have kept only those for which the radial velocity difference with the reference galaxy is $|\Delta V_r|<500$~km.s$^{-1}$.

This results in 6489 non-isolated galaxies (for which we have found one or more companions with these criteria), which means that about 50 \% of the UGC galaxies are isolated (see their list in \cite{isol}).


Then we have applied the reciprocity criterion to the galaxies with companions. This results in 2280 reciprocal pairs (where each galaxy of the pair is the closest to the other one, allowing to exclude multiplets like triplets, chains, etc.).

Finally, we start from this reciprocal pair list and we search the closest galaxy to the pair center in the whole UGC catalog with relative radial velocity difference $|\Delta V_r|<500$~km.s$^{-1}$. Then we keep only the pairs for which this closest galaxy shows a relative projected distance ratio $d_3/r_p > 2.5$. We are left with 1321 pairs satisfying the three criteria. 

Concerning the pair isolation, we note that the reciprocity criterion has excluded multiplets, chains, etc.. Now the question of the influence of the galaxy closest to the pair depends on the kind of study considered by the catalog user. Therefore, instead of choosing an arbitrary isolation criterion, we have determined for all 2280 reciprocal pairs the distance to the nearest galaxy of the UGC (up to a limit $\rho=10$). The corresponding distribution is shown in Fig.~\ref{rho3}. The value of $\rho$ is given in the catalog, allowing the user to extract any sub-catalog of isolated pairs according to his (her) own choice.

The criterion $\rho>2.5$ has selected about half of the reciprocal pairs (1246 among 2280). Then we find that 633 pairs (about $50\%$ of the pair catalog) are weakly isolated ($2.5<\rho<5$), 613 pairs ($50\%$ of the pair catalog) are truly isolated with $\rho>5$ and 307 pairs 
(25~\% of the catalog), are highly isolated ($\rho>10$), in agreement with the observed decrease as $\rho^{-2}$ of the $\rho$ probability distribution (see Fig.~\ref{rho3}). 

The additional isolation criterion using CGCG galaxies (with the isolation parameter $\rho_Z$) has decreased the sample of isolated pairs to a final number of 1005 ($16\%$ of UGC galaxies). The statistics remain essentially the same:   509 isolated  pairs have $\rho>5$ ($50 \%$ of the pairs, i.e. $8 \%$ of the UGC galaxies) and 273 are highly isolated with $\rho>10$ ($27 \%$ of the pairs, i.e. $4 \%$ of the UGC galaxies), where $\rho$ is still the isolation parameter refering to UGC galaxies.

An inner depletion is seen in this distribution for $\rho<3/2$ : this is a direct consequence of the reciprocity criterion, which implicitly contains an isolation criterion (see Fig.~\ref{cercles}). As a matter of fact, one can account accurately  for this depletion by simple geometrical considerations. Indeed, the reciprocal pairs are those for which the nearest UGC galaxy is on the part of the circle (C) of radius $\rho r_p$ centered on O (the red circle in Fig.~\ref{cercles}, in fact a narrow ring) which is outside the two circles of radius $r_p$ centered on A and B (the two members of the pair). This results in two symetrical arcs of (C) seen from O under an angle $\alpha$. If we admit that the density of UGC galaxies is constant along (C), the proportion $p$ of those reciprocal pairs is $p=\alpha/\pi$. This also implies that there is no reciprocal pairs for $\rho< \sqrt{3}/2 \approx 0.866$ (dashed circle in Fig.~\ref{cercles}), and that all the pairs are reciprocal if $\rho>3/2$.

It is easy to compute $\alpha$ by analytical geometry, namely one finds
\beq
\alpha= 2 \arctan  \l(  \frac{\rho^2-3/4}{\sqrt{1-(\rho^2-5/4)^2}} \r).
\label{eq4}
\eeq
This results in a proportion $p$ increasing quasi linearly between $0$ and $1$ when $\rho$ varies between $ \sqrt{3}/2$ and $3/2$, in excellent agreement with the observed distribution of reciprocal pairs (see Fig.~\ref{rho3}).

\section{Gathering of accurate radial velocities for members of the pairs of galaxies}
\label{sec4}

           One of the main purposes of our study of pairs of galaxies is to analyse the distribution of the actual (unprojected) velocity differences between their respective members. Such an analysis requires to use radial velocities $V_r$ of the pair members of our catalog as accurate as possible. We are not satisfied with the $V_r$ values given by Hyperleda, because they are a weighted average of all the measurements, some accurate and some other ones inaccurate. Even though the Hyperleda automatic method takes the uncertainties into account in the weighting, the presence of a very uncertain value is often enough to strongly bias the final velocity. Therefore, we prefer to select the most accurate ones and use only them. 
           
           In order to collect our accurate $V_r$ values, we have used the bibliography up-to-date in Hyperleda (by March 2014) and the NASA extragalactic database (NED, \cite{NED}) which gives for each galaxy listed the various $V_r$ measurements with their uncertainties (unfortunately not totally up-to-date).
           
           There are mainly two sources of $V_r$ measurements, namely those from the radio 21cm HI  line, and those from optical lines. The HI measurements are generally more accurate than the optical ones, except for the optical data coming from the Sloan Digital Sky Survey (SDSS-III), which have an accuracy better than 5 km s$^{-1}$, equivalent to HI ones.
           
          For the members of a pair of galaxies, the HI measurements present a risk of confusion when the angular distance between the members is smaller than the Half Power Beamwidth (HPBW) of the radiotelescope. Such a confusion is not necessarily visible in the HI spectra if the velocity difference between the members is small. In order not to introduce such an error, we have excluded the HI measurements relative to such pairs, taking only optical values if available; if not available for the two members, the pair has been rejected from our catalog. Note that the smallest HPBW of a single-dish radiotelescope is Arecibo's one (3.3 arcmin); thus only optical measurements have been taken into account for pairs with angular distances lower than 3.3 arcmin.
          
         Our main sources of accurate $V_r$ are the following:\\
         
\begin{itemize}
\item Optical measurements:

\hspace{0.5cm}(i) Aihara et al. \cite{Aihara2011}:   SDSSIII DR8  ;    $\varepsilon_{V_r}<5$ km s$^{-1}$.
 
\hspace{0.5cm}(ii) Huchra et al. \cite{Huchra2012}:   2MASS redshift survey   ; $\varepsilon_{V_r}<30$ km s$^{-1}$.\\

\item HI line measurements:

\hspace{0.5cm}(i) Springob et al. \cite{Springob2005}:   Homogeneous compilation of HI spectral  parameters for 9000 galaxies; median of $\varepsilon_{V_r}= 4$ km s$^{-1}$.

\hspace{0.5cm}(ii) Haynes et al. \cite{Haynes2011}: Arecibo ALFALFA survey; very accurate $V_r$,  $\varepsilon_{V_r}<2$~km~s$^{-1}$.
 
 \end{itemize}
 \vspace{0.5cm}

        After exclusion of pairs with possibly confused members, pairs with inaccurate $V_r$ ($\varepsilon_{V_r}> 100$ km s$^{-1}$) and two wrong pairs (\{UGC6539, UGC6545\} and \{UGC11913, UGC11919\}, identified by their very large velocity difference between the members), 
our final pair catalog lists 1246 isolated pairs of galaxies with $\Delta V_r <500$ km s$^{-1}$. The uncertainty on the radial velocity difference between the pair members is $\varepsilon_{\Delta V_r} =\sqrt{\varepsilon_1^2 +\varepsilon_2^2}$, $\varepsilon_1$ and $\varepsilon_2$ being the respective uncertainties on the radial velocities of the members.

\section{Final pair catalog: description}
\label{sec5}

The resulting pair catalog (Table 2) contains, for each pair (classified according to increasing UGC number of the first member): column (1), the UGC number of the first member; column (2), the UGC number of the second member; column (3), the radial velocity $V_1$ relative to the Sun of the first member, in km s$^{-1}$ ; column (4), the uncertainty $\varepsilon_{V_1}$  on $V_1$; column (5), the radial velocity $V_2$ relative to the Sun of the second member, in km s$^{-1}$; column (6), the uncertainty $\varepsilon_{V_2}$  on $V_2$; column (7), the absolute value $|\Delta V_r|$ of the radial velocity difference between the pair members, in km s$^{-1}$; column(8), the uncertainty $\varepsilon_{\Delta V}$ on $|\Delta V_r|$; column (9), the projected distance on the sky $r_p$ between the two pair members, in Mpc; column(10), the ratio $\rho=d_3/r_p$ between the projected distance $d_3$ of the nearest UGC galaxy to the center of the pair and the pair interdistance $r_p$ (a value $10.$ in the Table means that $\rho \geq 10$).


 \begin{table}
\caption{The UGC isolated galaxy pair catalog: example (first and last pairs). The full catalog is available  at CDS in VizieR \cite{vizier}.}
\vspace{3 mm}
\centering
\begin{tabular}{cccccccccc}
\hline \hline
 ${\rm UGC}_1$&${\rm UGC}_2$&$V_1$&$\varepsilon_{V_1}$&$V_2$&$\varepsilon_{V_2}$&$|\Delta V_r|$& $\varepsilon_{\Delta V}
$&$r_p$&$\rho$\\ 
 \hline

 18 & 23 & 7864 & 51 & 7975 & 4 & 111 & 51.2 & 0.222 & 10. \\
 20 & 12905 & 4153 & 22 & 4104 & 20 & 49 & 29.7 & 0.399 & 10. \\
 34 & 36 & 6137 & 2 & 6299 & 24 & 162 & 24.1 & 0.243 & 3.9\\
... & ... & ... &... & ... & ... & ... & ... & ... & ... \\
 12906 & 12919 & 5366 & 6 & 5486 & 4 & 120 & 7.2 & 0.391 & 3.1 \\
 12908 & 12911 & 4903 & 30 & 4768 & 7 & 135 & 30.8 & 0.018 &10. \\
 12914 & 12915 & 4371 & 8 & 4336 & 7 & 35 & 10.6 & 0.019 & 10.\\
 \hline
 \end{tabular}
 \label{table1}
\end{table}

\section{Properties of the pair catalog}
\label{sec6}

A detailed analysis of the catalog, in particular regarding statistical deprojections of the velocity difference and distance between members of the pairs, will be done in following papers. Presently we will present some general properties of the catalog.

\subsection{Distance distribution}

\begin{figure}[!ht]
\begin{center}
\includegraphics[width=12cm]{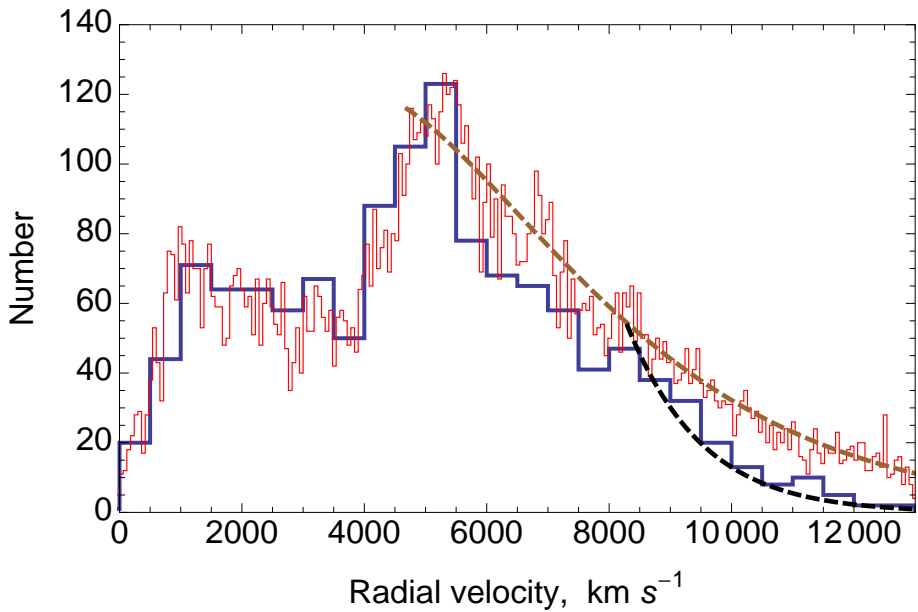}
\caption{{\small  Distribution of the pairs according to cosmological distance, measured by radial velocity (thick blue line). It is compared to the distribution of all galaxies in the UGC (thin red line) after normalization in the range $[ 0-5500]$ km s$^{-1}$.  The brown dashed curve is a fit by the expected exponential diameter function for the decrease of the number density of galaxies observed beyond $\approx 5500$ km s$^{-1}$. The black dashed curve is a fit of the expected loss of pairs, whose relative value quickly increases for increasing distances beyond $\approx 8000$  km s$^{-1}$ (see text).} }
\label{PairDistance}
\end{center}
\end{figure}

The distribution of cosmological radial velocities (and therefore of the distances) of the pairs is shown in Fig.~\ref{PairDistance} and compared to the distance distribution of the UGC catalog galaxies.

The agreement between the two distributions is globally satisfactory, suggesting that the pairs have been correctly drawn at random from the parent UGC distribution and that the rate of pairs up to $\approx 13000$ km s$^{-1}$ $\approx 200$ Mpc is almost constant.

However, one can notice that the decrease of the number of UGC galaxies at large distances ($V_r>6000$ km s$^{-1}$) due to the apparent diameter limit, is slower than that of the galaxy pairs. As shown in Sec.~\ref{diamfunct}, the decrease of the number of UGC galaxies is in excellent agreement with the Hudson and Lynden-Bell exponential diameter function for $V_r>7500$ km s$^{-1}$, at distances which are not affected by galaxy clustering. One can also account for the faster decrease of the pair distribution at those distances thanks to this diameter function. Indeed, the density $\rho(v)$ of UGC galaxies at a distance measured by the cosmological radial velocity $v$ is:
\beq
\rho(v) = \rho(0) \exp \l( -\frac{v}{v^*}  \r)
\eeq
from Eq.~(\ref{dNv}).

Let $\alpha$ be the proportion of visible UGC galaxies which belong to pairs at this distance. Only those pairs with a second galaxy visible at this distance can be identified as such, i.e., a proportion $p$ of real pairs given by
\beq
p= \exp \l( -\frac{v}{v^*}  \r),
\eeq
assuming that the galaxy pairing does not depend on the linear diameters (which also means that $\alpha$ is constant with the distance).

Thus the density of visible pairs at distance given by $v$ is:
\beq
\rho_p = \alpha \; \rho(0) \exp  \l( -2\frac{v}{v^*}  \r),
\eeq
resulting in 
\beq
dN(v)=\alpha \times 4 \pi f  \; \frac{\rho(0) }{H_0^3}  \; \exp  \l( -2\frac{v}{v^*}  \r) v^2 dv,
\eeq
where $f=0.42$ for the UGC (see Eq.~\ref{dNv}).

Figure~\ref{PairDistance} shows the corresponding curve $dN(v)$ for $v>7500$ km s$^{-1}$, in good agreement with the observed pair distribution.

\subsection{Uncertainties on velocity differences}

\begin{figure}[!ht]
\begin{center}
\includegraphics[width=12cm]{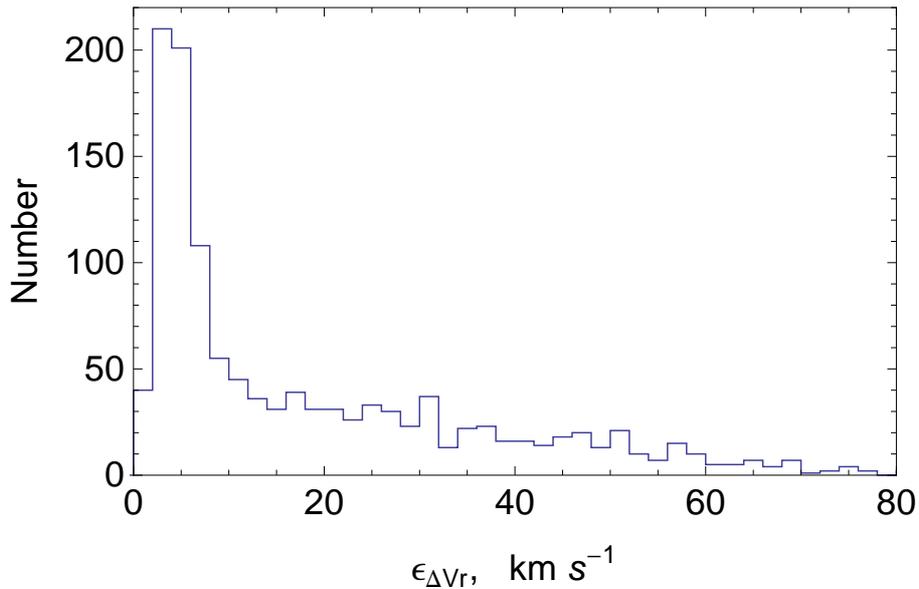}
\caption{{\small Histogram of the measurement uncertainties $\varepsilon_{\Delta V_r} =\sqrt{\varepsilon_1^2 + \varepsilon_2^2}$ on the difference between radial velocities of pair members. The bin width is 2 km s$^{-1}$. It shows a sharp peak of accurate velocity measurements at 4 km s$^{-1}$ (radio HI and SDSS measurements) and a tail of less accurate optical values.}}
\label{Verror}
\end{center}
\end{figure}

  Figure~\ref{Verror} shows the histogram of $\varepsilon_{\Delta V_r} $; note the large number of small $\varepsilon_{\Delta V_r} $-values. The median of $\varepsilon_{\Delta V_r} $ is $\varepsilon_{\rm med}= 10$ km s$^{-1}$, which is much lower than the median $\approx 200$ km s$^{-1}$ of the difference of radial velocities between the members of the pairs.

\subsection{Distribution of radial velocity differences between pair members}
\begin{figure}[!ht]
\begin{center}
\includegraphics[width=12cm]{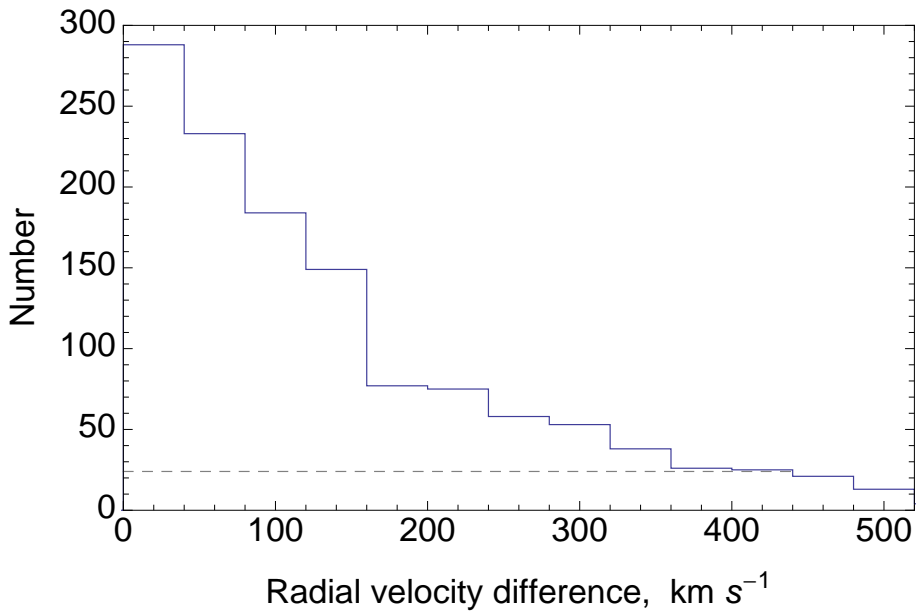}
\caption{{\small Histogram of the radial velocity differences between pair members, for a bin width of 40 km s$^{-1}$. We theoretically expect a monotonously decreasing distribution as a consequence of the projection effect (for randomly distributed orientations of velocity vectors): the obtained histogram supports this expectation. The dashed horizontal line is an estimate of the maximal contribution of cosmological false pairs, which becomes $100\%$ beyond $\approx 380$ km s$^{-1}$.}}
\label{DV40}
\end{center}
\end{figure}

 We give in Fig.~\ref{DV40} the histogram of the radial velocity differences between the members of the galaxy pairs, which is decreasing for increasing radial velocity as theoretically expected \cite{Tifft1977}. The small number of values beyond the catalog limit 500 km s$^{-1}$ comes from improved velocity values obtained once the catalog was completed.

We shall study in more detail this velocity distribution and its deprojection in a forthcoming paper.

\subsection{Distribution of interdistances between pair members}

\begin{figure}[!ht]
\begin{center}
\includegraphics[width=12cm]{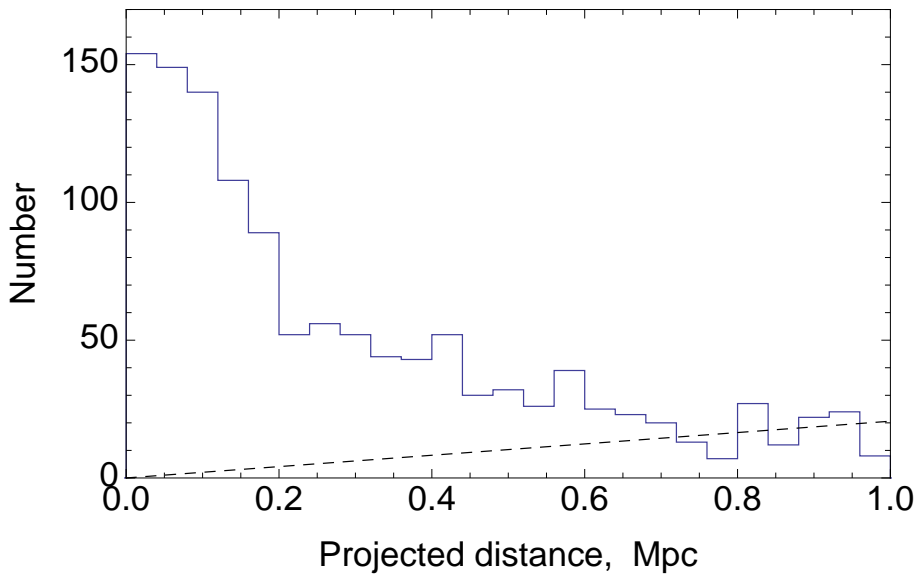}
\caption{{\small Histogram of the projected distances between pair members, in bins of 0.04 Mpc. The dashed line is an estimate of the maximal contribution of cosmological false pairs. }}
\label{dist}
\end{center}
\end{figure}

We give in Fig.~\ref{dist} the distribution of projected distances between pair members in our catalog. Two well defined populations clearly appear, with two different distributions of number density, one with $d<0.2$ Mpc, the other one with $0.2<d<0.75$ Mpc. The density almost vanishes at this value $0.75$ Mpc. Some pairs remain between $d=0.75$ Mpc and the cut-off of our catalog 1 Mpc, which are compatible with being false (cosmological) ``pairs". The mean projected distance of the whole sample is $<d> = 0.29$ Mpc.

\subsection{False (cosmological) ``pairs"}

Some of the pairs are expected to be the result of mere projection effects, their members lying actually at a large relative distance. Indeed, the radial velocity criterion $|\Delta V_r|<500$ km s$^{-1}$ was chosen in order to include very tight pairs, which have large intervelocities due to Kepler's third law. However such a limit for the velocity difference corresponds also to a large cosmological distance,  $d_c=DV/H_0=7$ Mpc (taking a Hubble constant $H_0=70$ km s$^{-1}$ Mpc$^{-1}$). One exceeds our interdistance criterion $r_p<1$ Mpc from $|\Delta V_r|=70$ km s$^{-1}$.

We have attempted to estimate this contamination by cosmological false ``pairs" by using its expected dependence on radial velocity and projected interdistance. The volume defined by our criteria is almost a radial cylinder along the line-of-sight, so that the number of false pairs depends linearly on $|\Delta V_r|$. Therefore the corresponding rate in the interval $[|\Delta V_r|,|\Delta V_r|+d|\Delta V_r|]$ is constant (see Fig.~\ref{DV40}). Concerning the projected interdistance, the volume depends on it quadratically, so that the expected dependence on $r_p$ in the histogram is linear, $N \propto r_p$. We have performed numerical simulations which have confirmed these expected dependences.

The observed distribution of $|\Delta V_r|$ supports this expectation, since it becomes almost flat beyond $\approx380$ km s$^{-1}$ (then falls down around the limit of the catalog 500 km s$^{-1}$). The value  $380$ km s$^{-1}$ is a reasonable limit for the maximal velocity difference between members of real pairs. Identifying this flat tail (containing $\approx 60$ pairs) to the expected flat cosmological pair distribution yields an upper limit to the false pair contribution.  Since the observed constant rate in the range $\approx 380-500$ km s$^{-1}$ is $\approx 20$ pairs by bins of 40 km s$^{-1}$ (see Fig.~\ref{DV40}), we estimate by this way the maximal total number of false pairs to be $\approx 250$, i.e. $\approx 25$\% of the pair catalog. This rather large number is a consequence of our voluntary choice to (i) take a large limit for our radial velocity criterion $|\Delta V_r|<500$ km s$^{-1}$, then (ii) correct for cosmological contamination. We have preferred such a strategy rather than to risk having missing pairs.

Concerning the projected interdistances, the observed distribution is also compatible with an expected linear false pair contribution (see Fig.~\ref{dist}). Indeed, the histogram shows a clear change of behavior, decreasing up to a minimum at $\approx 0.75$ Mpc, then increasing again up to the catalog limit at 1 Mpc. By considering all pairs beyond this minimum as false cosmological pairs, we find another estimate of the maximal number of these pairs equal to 210, in reasonable agreement with our previous determination from velocities (250). Finally, the number of false pairs in the reduced catalog $|DV|<380$ km s$^{-1}$ and $r_p<0.75$ (containing 860 pairs) is at most $90 \pm 20$, so that its rate has fallen down to less than only 10\%.

\section{Conclusion}
\label{sec7}

In this paper, we have carried out the construction of a sample of isolated galaxy pairs, using the Uppsala General Catalog of Galaxies (UGC). For this purpose, we have selected accurate quantitative criteria to define isolated galaxy pairs, namely: 1) Low radial velocity difference between the pair members: $|\Delta V_r|<500$ km s$^{-1}$; 2) Small projected distance between the pair members: $r_p<1$ Mpc; 3) Reciprocity, allowing to exclude multiplets; 4) Relative isolation: the catalog lists the ratio $\rho$ between the distance of the closest UGC galaxy (having a velocity difference with the pair smaller than 500 km s$^{-1}$) and the pair members interdistance for $\rho>2.5$, thus allowing any choice of isolation criterion beyond this value;
5) Additional isolation condition for excluding pairs having  close-by CGCG galaxies.

Our final catalog contains 1005 galaxy pairs with $\rho>2.5$, of which 509 are isolated with $\rho>5$; this sub-sample contains $ \approx 50\%$ of the pairs, i.e. $\approx 10\%$ of the UGC galaxies. This rate is the same as obtained by Karachentsev and Makarov \cite{Karachentsev2008} for isolated pairs (using different criteria). Finally, 273 pairs are highly isolated with $\rho>10$ ($\approx 25\%$ of the pairs, i.e. $\approx 5\%$ of the UGC galaxies), which is the same rate as in the pioneering 1972 Karachentsev's catalog \cite{Karachentsev1972}.

      Applying those criteria leads to a sample of 1005 galaxy pairs, for the members of which we have collected today's most accurate radial velocities. The distribution of the pairs versus their cosmological distance is nearly identical to that of the UGC galaxies, excluding a possible distance bias. The proportion of ``false" cosmological pairs, i.e. galaxies which seem close to each other due to projection effects, is estimated to be less than 25\% in the full catalog, and less than 10\% in the reduced catalog $|\Delta V_r|<380$ km s$^{-1}$ and $r_p<0.75$.
      
      Thus we have at disposal a quite fair and large sample, which allows significant statistical studies. This pair catalog construction is a first step before its extension in a future work using the now available large databases (which have been used here for the accurate redshift search).
      
      In a forthcoming paper, we intend to perform a dynamical study of those galaxy pairs, in order to obtain the masses of their members, their mass-luminosity ratios, and possibly to check the presence of massive haloes and of dark matter in them. Such a work needs a statistical derivation of the actual velocity differences between the members and of their actual interdistances from their known measured projections, a task which will be developped and used in the future analysis of our galaxy pairs catalog.\\

{\bf Acknowledgements.}
We acknowledge the use of the HyperLeda database\\ 
(http://leda.univ-lyon1.fr). 
This research has made use of the NASA/IPAC Extragalactic Database (NED) which is operated by the Jet Propulsion Laboratory, California Institute of Technology, under contract with the National Aeronautics and Space Administration.\\
We acknowledge very helpful remarks from an anonymous referee concerning the isolation criterion, which have allowed to improve the catalog.\\


\end{document}